# SHOCK INDUCED DAMAGE MECHANISM OF PERINEURONAL NET

KAH Al Mahmud, Fuad Hasan, Md Ishak Khan, Ashfaq Adnan[1]

Mechanical and Aerospace Engineering, the University of Texas at Arlington

**Abstract:**

ECM components, such as the Perineuronal net (PNN), one of the most prevalent parts surrounding the neuronal cell. PNN is a protective net-like structure regulating neuronal activity such as neurotransmission, charge balance and generates an action potential. Shock induced damage of this essential component may cause neuronal cell death and potentially leads to CTE, AD diseases, PTSD, etc. The shock generated possibly during an accident, improvised devie explosion or collision between NFL players may lead to damage to this safety net. The goal is to investigate the mechanics of PNN under shock wave. To understand the mechanics of PNN, mechanical properties of different PNN components such as glycan, GAG, and protein need to be evaluated. In this study, we evaluated the mechanical strength of PNN molecules and the interfacial strength between the components of PNN. Afterward, we have assessed the PNN molecules' damage efficiency at various conditions such as shock speed, preexisting bubble, and boundary conditions. The secondary structure altercation of the protein molecules of the PNN has been analyzed to evaluate damage intensity under varying shock loading. At higher shock speed, damage intensity is more elevated, and hyaluronan is most likely to break at the rigid junction. The primary structure of the protein molecules is most unlikely to fail. Instead, the molecules' secondary bonds will be altered. Our study suggests that the number of hydrogen bonds during the shock wave propagation decreased.

[1] Corresponding Author: Dr. Ashfaq Adnan
Email: aadnan@uta.edu

## 1. Introduction:

Concussion, sub concussions, and exposures to the shock waves from the explosive blast can cause mild traumatic brain injury (mTBI) [1]. A typical blast-induced shock wave profile exhibits a sudden increase in pressure, often referred to as overpressure, followed by a low magnitude longer duration negative pressure tail [2]. For example, the primary ingredient in RDX (Royal Demolition eXplosive) can generate an initial overpressure of over 27 GPa [3]. The long-range negative pressure tail causes damage to the Extracellular Matrix (ECM) and neuronal cells by forming micro cavitation [2], [4], and mechanical fracture of different biomolecules [5], [6][7][8]. The overpressure generates a compressive load, which may cause shear fracture of biomolecules. Perineuronal net (PNN) is a critical ECM component. PNN gives neuroprotection to the neuronal cells by forming a safety net. Apart from that, it regulates synaptic plasticity and protects neuron cells from Oxidative stress. Therefore, the investigation of the behavior of PNN under shock loading is essential.

PNN is an interconnected net-like structure that consists of three significant biomolecules, such as Lectican, Tenascin-R, and Hyaluronan. The Lectican family of the chondroitin sulfate proteoglycans (CSPGs) is the most prominent in PNN of the Central Nervous System (CNS). Lectican consists of a Core Protein (CP) with covalently connected negatively charged Glycosaminoglycans (GAGs) side chains [9]. Tenascins are globular proteins that bind to the C-terminal domain of CP. Hyaluronan (HA) is an unsulfated GAG synthesized at the cell surface by the enzymes known as HAS. This membrane-bound enzyme emanates through the plasma membrane into the extracellular space [10]. Hyaluronan binds to the N terminal of other ECM protein molecules such as CP and various link proteins (LP) [11].

The absence of PNN in the neuron may cause a severe problem. Sometimes aberrations in their molecular structure also affect the functionality of the neuron. For instance, researchers have found that a reduction in normal PNN densities in the brain areas often affects cognitive functions in subjects with Alzheimer disease (AD) [12]. PNN loss in AD may contribute to altered excitatory/inhibitory balance, synaptic loss, and increased susceptibility to oxidative stress [9]. Some studies have also suggested that TBI is associated with an earlier onset of AD [13]. It is undeniable that structural alteration or loss of PNN in the brain plays critical role in neurodegenerations and cognitive functions.

The shock wave can break PNN in the presence of nanobubbles. Studies have shown that shock velocity and cavitation bubble size affect the fracture potential of PNN [5]. The higher shock speed and bigger bubble size are more damaging. Several studies have been conducted on the shock wave effect on the transport phenomena and the deformation strain of the lipid bilayer [14]–[16]. However, the deformation mechanism of the PNN component due to the shock wave has not been studied yet. The morphological degradation of PNN due to shock waves causes several diseases and alter the action potential by damaging the synapse. Propagation of the shock waves through the brain tissue damages the primary or secondary bonded structure of several PNN molecules. PNN molecules include proteoglycan, tenascin-R, link protein, and hyaluronan. Among them, hyaluronan, which is the backbone of PNN more prone to damage [5].
In this study, different junctions of PNN, which are noncovalently bonded, are studied to assess their relative strength. There are three interfaces among various components of PNN, such as 1. link protein (LP) Core Protein (CP) of proteoglycan (PG), 2. Tenascin-R (TR) and CP, 3. LP and Hyaluronan (HA). The relative strength of the junctions has been evaluated to figure out the weakest link among the molecular interfaces of the PNN. Afterward, the damaging efficiency of the shock wave propagation through the PNN model has been studied.

## 2. Methodology:

### 2.1 Modeling of PNN Structure

Perineuronal-net (PNN) is a protective ECM component that surrounds the neuronal cell (*Figure 1*a). The basic building block of PNN includes Hyaluronic acid (HA), Proteoglycans (PG), Tenascin-R (TR), and Link Protein (LP) (*Figure 1* b). Proteoglycan consists of Core Protein (CP) and Glycosaminoglycan (GAG) chains connected to the core protein via a glycosidic covalent bond. GAG chains of the CP are negatively charged. Therefore these GAG side chains help to balance the charge distribution of neuronal cells. These small GAG chains have less contribution to the mechanical stiffness of the PNN; thus, in this study, GAG chains are omitted. In this study, the PNN structure consists of HA, TR, CP, and LP. The TR and CP are connected by non-covalent bonding, HA, and CP connected by non-covalent or glycosidic covalent bonds mediated by LP. Docking protocol is used to model the most energetically favorable protein complex. For the protein-protein docking, ClusPro online server has been used [17] to perform the molecular docking. ClusPro introduced PIPER, an FFT based docking program. It uses a pairwise interaction potential as part of its scoring function E, where E is

$$E = E_{attr} + w_1 E_{rep} + w_2 E_{elec} + w_3 E_{pair} \qquad (2\text{-}1)$$

$E_{attr}$ and $E_{rep}$ denote the attractive and repulsive contributions to the van der Waals interaction energy $E_{vdw}$, $E_{elec}$ is an electrostatic energy term, and $E_{pair}$ represents the desolvation contributions. The coefficients w1, w2, and w3 specify the related terms' weights and are optimally selected for different docking problems [18].

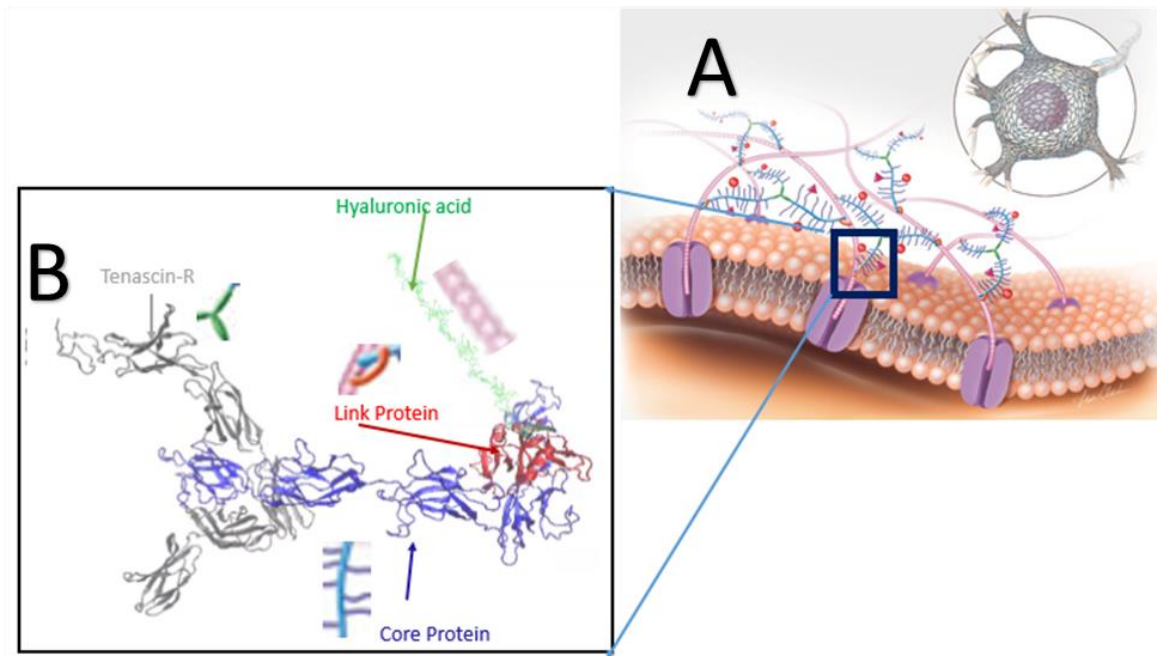

Figure 1 (a) Schematic illustration of PNN structure. The overall macromolecular structure of the PNN is obtained by the specific arrangement and binding of the components [28]. A major component of PNN is chondroitin sulfate proteoglycans (CSPGs) that include a core protein (CP) (blue) and several sugar chains (purple). Structurally, the CPs are bound to hyaluronic acid (HA) (pink). A set of link proteins (LP) (orange) are also present in PNN to stabilize the interaction between HA and CPs. Sema3A (pink pyramids) and Otx2 (red balls) are linked with the sugar chains of the CSPGs. Tenascin-Rs (green) are proteins in the PNN that are cross-linked with the CSPGs. (b) Docked PNN model structure.

At first, LP and CP are docked, then the LP-CP protein complex is further anchored with TR to get the final protein complex (LP, CP, and TR) of the PNN. Finally, a hyaluronic acid chain is attached with LP by glycosidic covalent bond using CHARMM-GUI Glycan Reader and Modeler module [19]. The HA chain consists of 15 repeated dimer of β-N-Acetylglucosamine and β-D-glucuronic acid linked via alternating β-(1→4) and β-(1→3) glycosidic bonds. The PNN unit is shown in Figure 1b; it represents the inset portion of Figure 1a.

## 2.2 Interfacial Strength of PNN components

ClusPro server has been used for the protein-protein docking (*Figure 2* (B and C)), and LP-HA (*Figure 2* A) complex has been taken from the original PDB structure of the HA binding domain of murine CD44 from the RCSB protein data bank (PDB ID: 2JCQ) [20]. ClusPro generated 100 energy minimized structures; only the most energy minimized configuration is taken for this study. The minimum energy configuration has been shown in *Figure 2* (B and C). *Figure 2* D shows the top 4 energy minimized structure of the CP-LP complex.

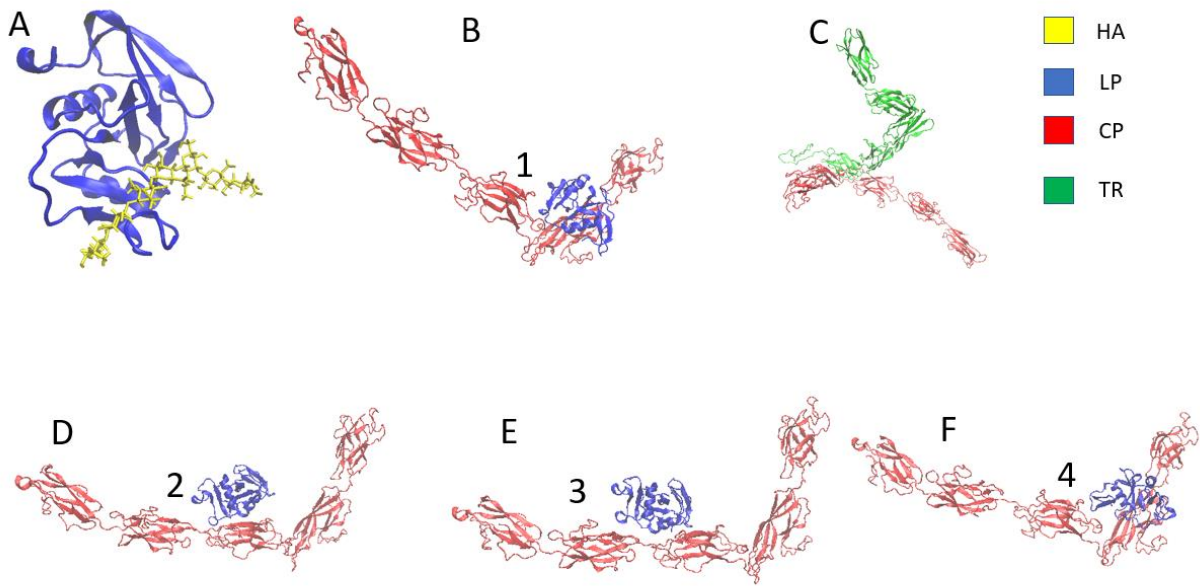

Figure 2 Docked Structure (A) HA-LP complex (B) LP-CP complex and (C) CP-TR complex (B and D-F) Top 4 energy minimized structure of the LP-CP complex (Proteins are represented in Newcartoon representation and Hyaluronan is represented in bonded representation)

Three complex structures have been tested for the relative interfacial strength calculation using Steered Molecular Dynamic (SMD) approach in GROMACS 5.0 simulation platform [21]. The LP molecule is fixed of the LP-HA protein-ligand complex and pulled the HA at a constant speed, while for the LP-CP and CP-TR protein-protein complex, LP and TR have been pulled at a steady rate. The molecules have been pulled from their center of mass.

**2.3 Mechanical property Evaluation of PNN components:**

The mechanical strength of the components is vital to evaluate the underlying mechanics of PNN under shock wave. Since one of the significant functions of PNN is to give neuronal protection from mechanical damage, the interfacial strength of the protein complex and the individual strength of the components needs to be evaluated. In this study, the CP and HA's mechanical properties have been assessed using the CHARMM36 and ReaxFF force field, respectively. Due to the CP and TR's structural conformational similarity, it can be assumed that CP and TR's deformation profile will be similar. Because of the globular secondary structure, the primary covalent bond break is very unlikely for the protein molecules. Instead, the applied force will cause secondary structure failure. The secondary structure of CP is globular, where each globule is connected by chain structure (*Figure 3*A). The whole structure contains a single chain. In comparison, the HA chain has ten dimers.

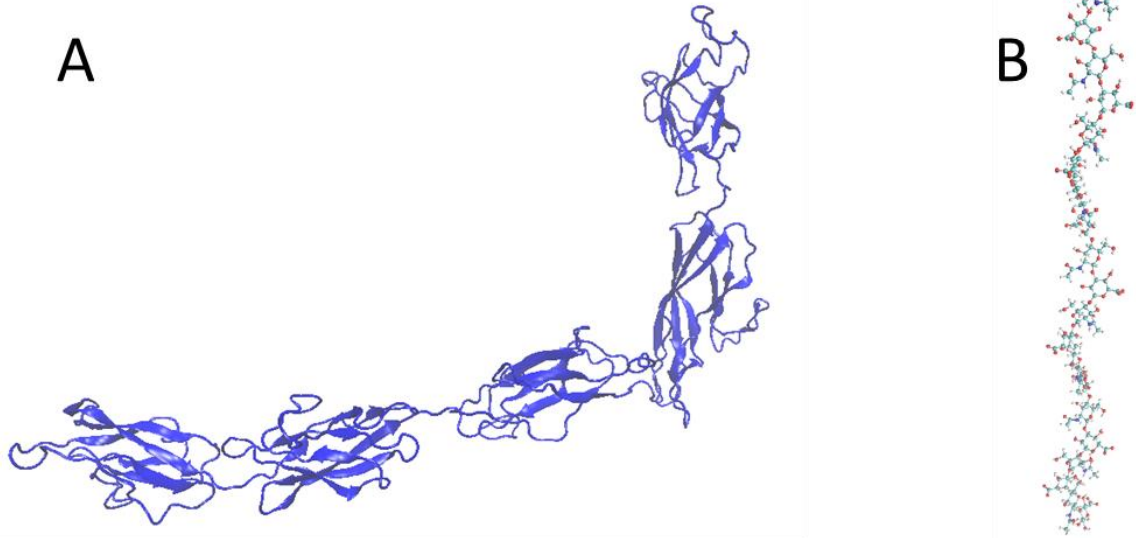

Figure 3 Structure of (A) CP (New cartoon representation) and (B) HA (all-atom representation)

The SMD approach evaluates the mechanical Strength of CP and HA. Few atoms at the end of the molecules have been pulled at a constant velocity of 1000 $ms^{-1}$ While the other end was fixed at the initial position. The temperature was maintained constant at 310K in all the simulations. In the beginning, the system is energy minimized by using the shaking algorithm. After that, the energy minimized structure is equilibrated at the NPT ensemble, where temperature and pressure remain constant, afterward pulling simulation is conducted in the NVT ensemble. In NVT, volume and temperature remain constant.

## 2.4 Shock simulation

The PNN structure of Figure 1b is used to conduct the shock simulation. The PNN model is solvated with TIP3P water and ions (0.1M NaCl) using CHARMM-GUI Glycan Reader and Modeler [19] module. The box size is 26.2 × 26.2 × 26.2 nm³, with full of water and PNN molecules. The X-direction was a shock wave propagation direction. The shock wave was formed from a negative direction and propagated to the positive X direction. Both the end of X direction are opened up to create a vacuum space so that it is possible to restrict the shock flow to the opposite end because the periodic boundary condition is applied along the shock direction. There are different ways of generating shock waves, one of the most common methods is "moving piston" [22][23][24], and "reflecting boundary" is another popular method widely used [25][26][5] [88]. The piston-driven shock has several advantages over the reflecting boundary method. The initial number of particles in a cell remains relatively constant throughout the simulation until the shock wave nears. Therefore density remains constant in the upstream region. Secondly, the simulation of the piston-driven shock wave closely resembles the corresponding physical experiment. Even though assumptions are made on the nature of particles' interactions and between particles and computational boundaries, the model can simulate otherwise difficult experiments [88]. To initiate the shock few layers of water molecules from the right end of the X axis are made rigid and pushed at a constant velocity for a certain distance and then the piston motion is halted. Piston are moved at a 1 km/s 2.5km/s and 4km/s velocity for 30Å distance. Density distribution of the simulation box shows that the region where PNN network molecules are present shows a 2% reduced (0.96 gm/cc) density compared to water density (0.98 gm/cc) at 310 K.

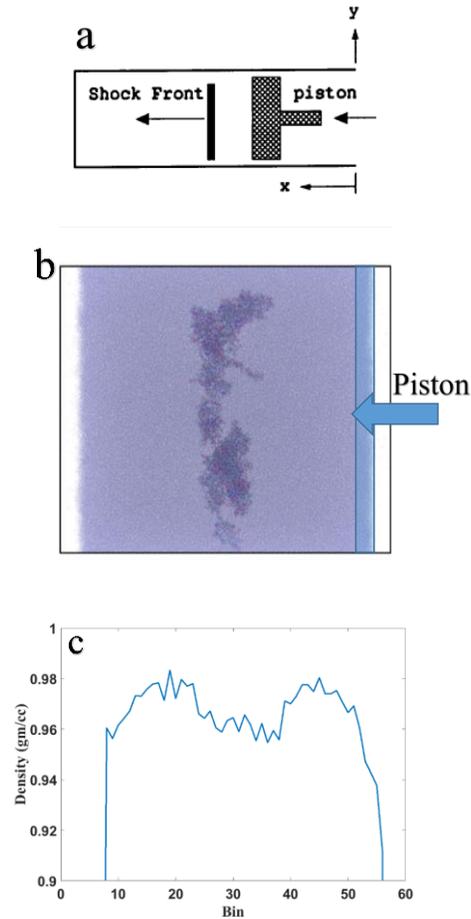

Figure 4 The Simulation box for shock propagation (a) Schematic illustration of the shock simulation setup [88] (b) Snapshot of the shock simulation box by Ovito visualization tool (c) Density profile along the shock direction.

## 3. Results and discussions:

### 3.1 Mechanical Strength of PNN components

The mechanical strength of the ECM components, such as core protein (CP) and hyaluronic acid (HA), is measured by SMD simulation. The CP is a very long protein coil chain that forms a secondary structure known as alpha-helix and beta-sheet. These secondary structures are strong and bonded by intrachain or interchain hydrogen bond, electrostatic, and Van der Waals interaction. During the pulling simulation, the secondary system breaks, which is mostly a non-covalent electrostatic bond. Covalent bond breakage for the protein component is very rare. Therefore it is wise to use a non-reactive charmm36 force field for CP. The non-reactive Charmm36 force field is advantageous over the reactive because it is widely used for biomolecules, and simulation is high-speed compared to the ReaxFF reactive force field.

In *Figure 5*, the CP is stretched 95% of its initial length, and the maximum pulling force is only 400 pN. In contrast, the maximum pulling force for hyaluronic acid at 40% stretch is 4500 pN, more than ten times the maximum CP stretched force. The HA covalent bond breaks at above ~45% strain. From the force-displacement curve of HA (*Figure 5*b), it can be found that at the toe region, the secondary bond stretched, and after that, around 20% strain covalent bond stretching starts and finally failed at 45% strain. The stiffness constant at the toe region is ~20pN/ Å. In contrast, at the covalent bond stretching region, it is ~160 pN/ Å, stiffness constant is eight times higher at the covalent bond stretching part than the toe region.

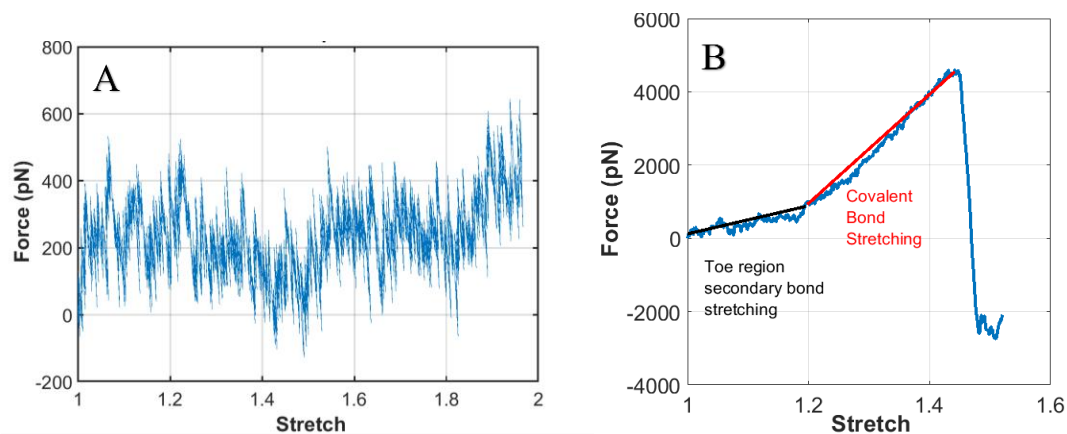

*Figure 5* Mechanical Strength of PNN components at 1 $kms^{-1}$ pulling speed (A) Core protein (CP) (B) Hyaluronan (HA).

**3.2 Interfacial strength:**

In the PNN network, three different interfaces exist, such as CP-LP, LP-HA, and CP-TR. Interfacial strength plays a significant role in the mechanics of PNN under shock wave. LP, HA, and TR from the three interfaces are pulled, whereas other molecules (CP, LP, and CP) of the pairs are kept fixed at their initial position. The mass of CP>TR>LP>HA for the PNN model. Although in reality, the molecular mass of HA is maximum because of its very long chain. *Figure 6* shows the relative interfacial strength of three different interfaces. It has been found that HA-LP has the lowest strength, and CP-TR is the highest. The CP-TR bonds never failed during the simulation. Instead, the TR molecules unfolded. The interfacial strength for the pair of molecules considered is between 1100 pN to 1500 pN. This strength is well below the fracture strength of the covalent bond. The LP-HA interface fails at around 1100 pN force, whereas the LP-CP fails at 1400 pN force. The CP-TR did not fail during the simulation period.

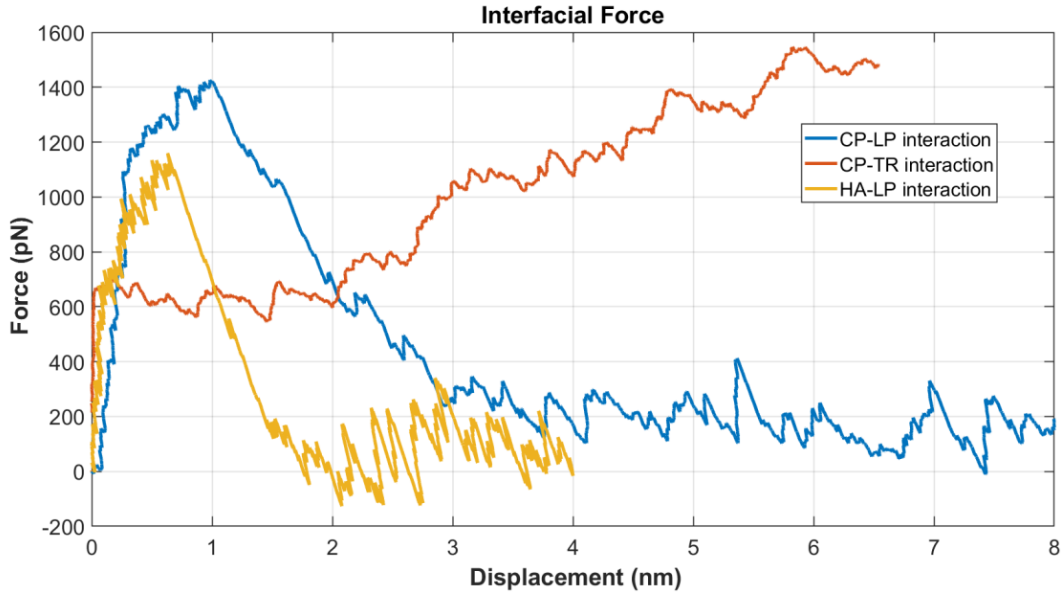

*Figure 6* Interfacial Strength of PNN components at 1 $kms^{-1}$ pulling speed

## 3.3 Shock simulation:

In this work, a piston is used to initiate the shock. A piston is moved up to 30 Å towards the positive X-axis at 1 km/s, 2.5 km/s, and 4km/s speed to initiate different shock speed. The inset plot of Figure 7a shows the corresponding shock velocity at different piston speeds.

### 3.3.1 Effect of Shock Speed

The shock wavefront is highly densified, which is called the overpressure region. After the overpressure region, there is a sharp decrease of density, density profile with time along the shock propagation direction shown in *Figure 7*b. The simulation box is divided into 61 bins along the shock propagation direction, and each bin is 5Å in size. The average properties, such as velocity, density, and pressure, are calculated for each bin's particles. The maximum density decays as the shock propagates, the decay rate at higher piston speed is much higher (*Figure 7*a). Peak pressure at different bin locations in *Figure 7*c shows that maximum decay is observed at 4 km/s piston speed, and 1km/s, minimum pressure decay along the shock direction. The decay rate at the middle (bin 25 to 30) of the simulation box is higher for 4 km/s and 2.5km/s piston speed. However, no significant change of decay constant has been observed for 1km/s piston speed. PNN molecules in the middle of the simulation box may impede water molecules' motion as the shock propagates. Thus the peak pressure dropping rate is highest in this region compared to the other areas.   However, the penetration of water molecules has not been hindered by the presence of PNN molecules at 1km/s. Therefore no significant change in the peak pressure observed in this region; implies that water molecule penetration efficiency depends on the pressure impulse. At higher pressure, the impulse penetration rate is lower [15].

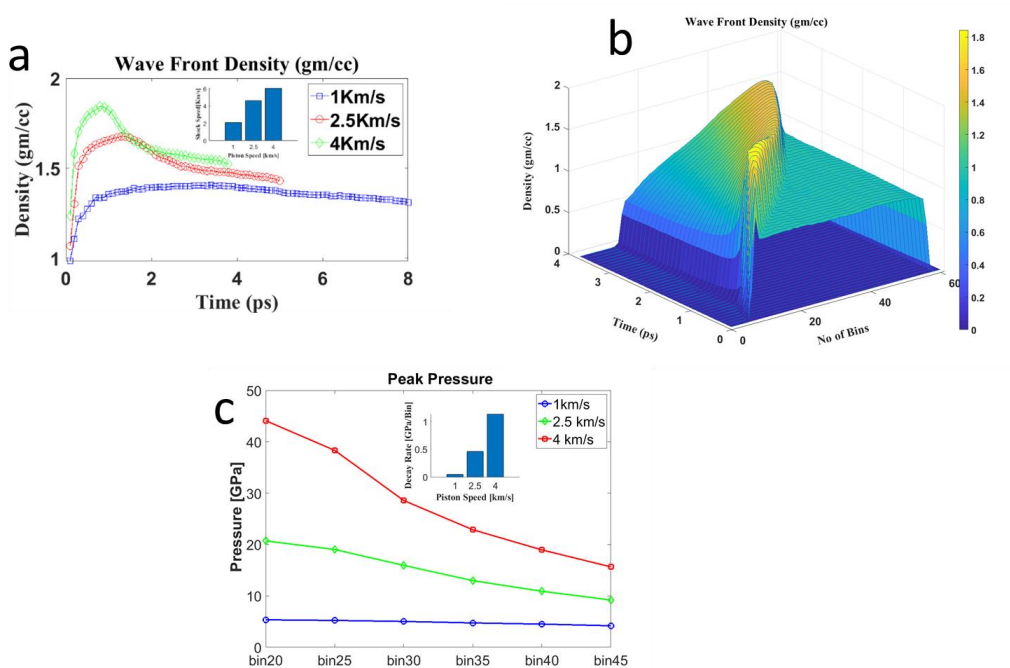

Figure 7 (a) Shock wavefront density at different piston speeds. The inset plot shows the piston speed's corresponding shock velocity (b) Density distribution during the shock propagation at different location and time for 4km/s piston speed (c) Peak pressure at different bin location along the z-axis during the shock propagation at different shock speed. The inset plot shows the pressure decay rate at different shock speed (each bin is 5Å along X direction)

While the shock propagates, different molecules experience different levels of pressure. *Figure 8*a and b show the pressure profile in CP and HA. These two molecules are considered because other molecules are protein. They are representative of the structural and bonding conformation of the CP molecule. Therefore, it is reasonable to assume that LP and TR will experience similar pressure as CP experiences. Peak overpressure depends on the shock speed, as the shock propagates CP experience 17 GPa, 9GPa, and 3.5 GPa compressive stress, while HA experience 12.5 GPa, 6 GPa, and 2GPa at 4, 2.5, and 1 Km/s piston speed. It is quite interesting to note that CP does not experience any tensile stress in almost every case, while HA experiences tensile peak pressure of around -5GPa, -3GPa, and -1GPa. This study does not confirm if HA will break or not. However, it is reasonable to approximate the cross-sectional area of HA around 50 to 100 Å² failure stress will be around 9 GPa to 4.5GPa. This value corresponds to the failure force of the HA in *Figure 5*b. The approximate value of fracture stress suggests that at 4Km/s piston speed corresponds to 6km/s shock speed, and HA will most likely break. With the fact that here average pressure of all atoms of the HA molecule is considered, there is a possibility that localized pressure will surpass the fracture stress at rigid junction even at lower shock speed.

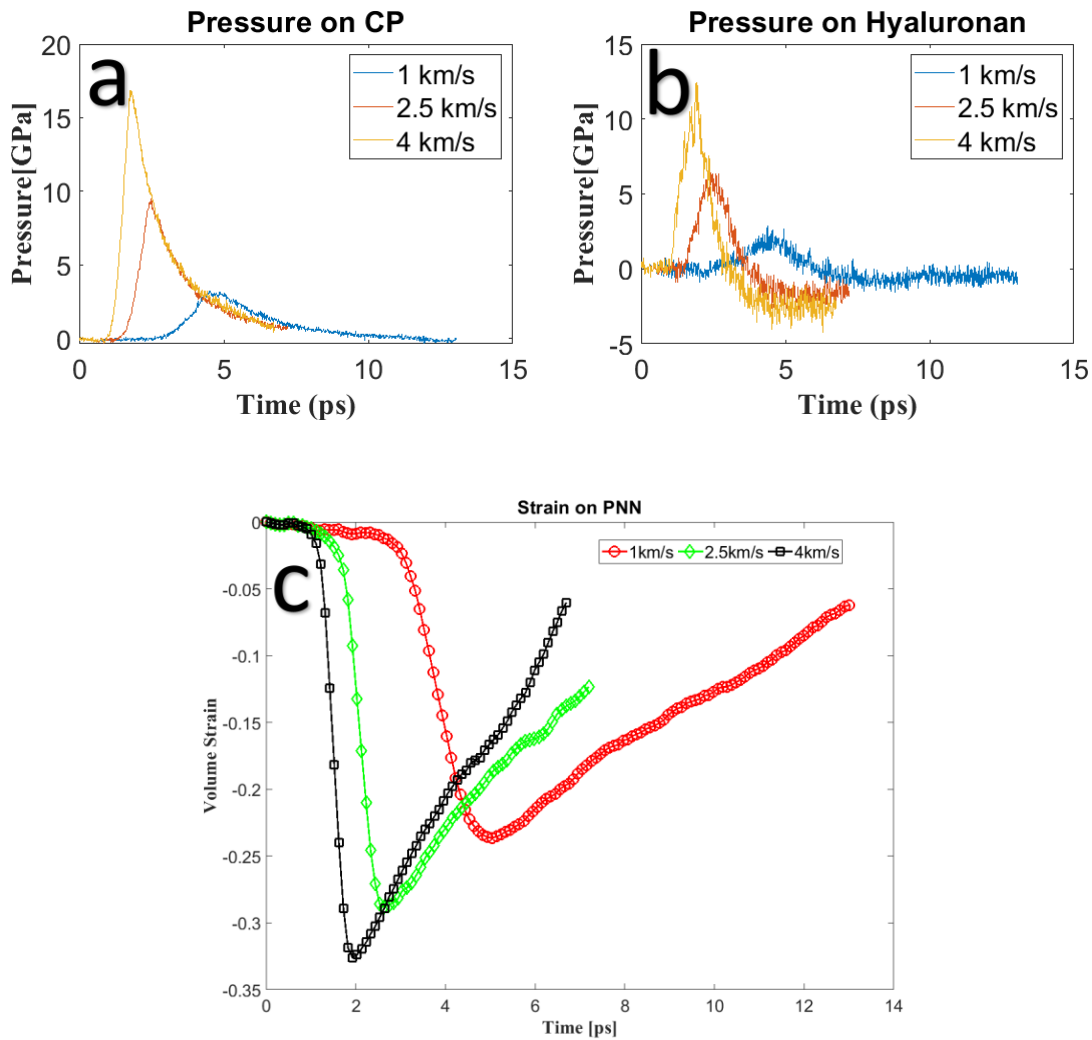

*Figure 8* Pressure on PNN components at different piston speed (a) Pressure on Core Protein (CP)(b) Pressure on Hyaluronan (HA) (c) Volumetric Strain on PNN at different shock speed.

### 3.3.2 Effect of cavitation bubble

Research work has been conducted on the effect of the bubble on the damage mechanics of biomolecules under shock loading [5][27][25]. The collapse of the bubble can be symmetric or asymmetric depending on the ratio of collapsing time $t_c$ and shock passing time $t_{sp}$, if the ration ($\frac{t_c}{t_{sp}}$) is higher than 1 the bubble will collapse symmetrically; otherwise, an asymmetric collapse will initiate. The ratio is related to the bulk modulus ($B_L$) and peak pressure difference ($\triangle p$) (Eqn ((3-1)). The bulk modulus of water is around 2.2 GPa, which means the cavitation bubble in water can only be asymmetrically collapsed by a shockwave having more than 2.2 GPa post-shock pressure or near that scale[25].

$$\frac{t_c}{t_{sp}} = \sqrt{\frac{B_L}{\triangle p}} \tag{3-1}$$

In this case, the peak pressure is way higher than 2.2 GPa; it is reasonable to assume that the bubble will collapse asymmetrically. The bubbles' asymmetric collapse forms a water-jet that can reach further away from the cavitation epicenter and cause more damage. It has been investigated the pressure profile of the PNN components in the presence of a bubble of 8nm diameter and found an insignificant change of overpressure (*Figure 9*). However, the overpressure on the CP molecule's bubble projected zone is higher than the without bubble model for 4km/s piston speed (*Figure 9*b). The portion of the CP molecule on the bubble projected zone is shown in *Figure 9*c.

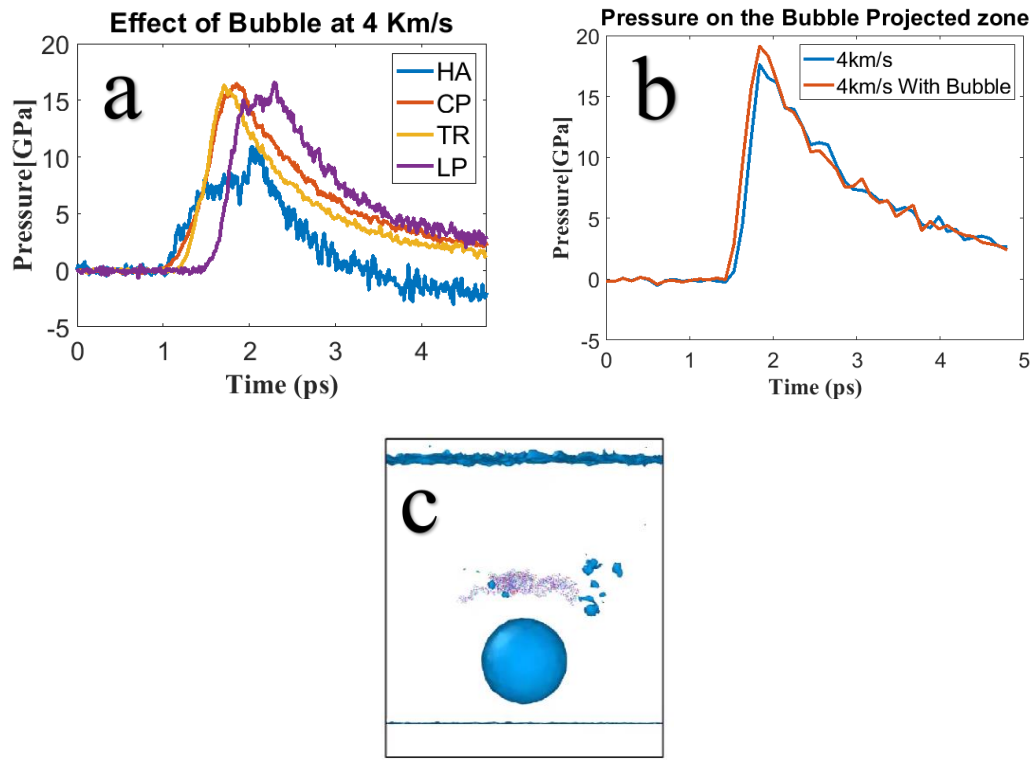

*Figure 9* Bubble induced shock propagation (a) Pressure on different PNN components (b) Pressure on Bubble projected area (c) bubble projected area of the PNN

It is quite interesting to note that although the maximum average overpressure of the components falls in the presence of a bubble. These projected overpressure increases give us the impression that the presence of a bubble enhances localized damage.

### 3.3.3 Effect of boundary conditions

The PNN components are interlinked by covalent and non-covalent (electrostatic and Van der Waals bonding). There will be a difference in the acceleration profile of the components.

Therefore it is reasonable to assume that one part of the molecule may experience 15 GPa, while other parts still at the atmospheric pressure. The links are often considered rigid as compared to the molecule itself. It has been observed that only the HA molecule experience negative tensile stress due to fixing one end. Although two atoms at the two end of TR are fixed, TR's average pressure does not change significantly (*Figure 10*a and b). HA mostly experiences tensile stress due to its chain conformation; the proteins are globular structure. It has been found that the fixed end of HA experienced higher tensile pressure than the middle portion, and the part connected to the LP (*Figure 10*c).

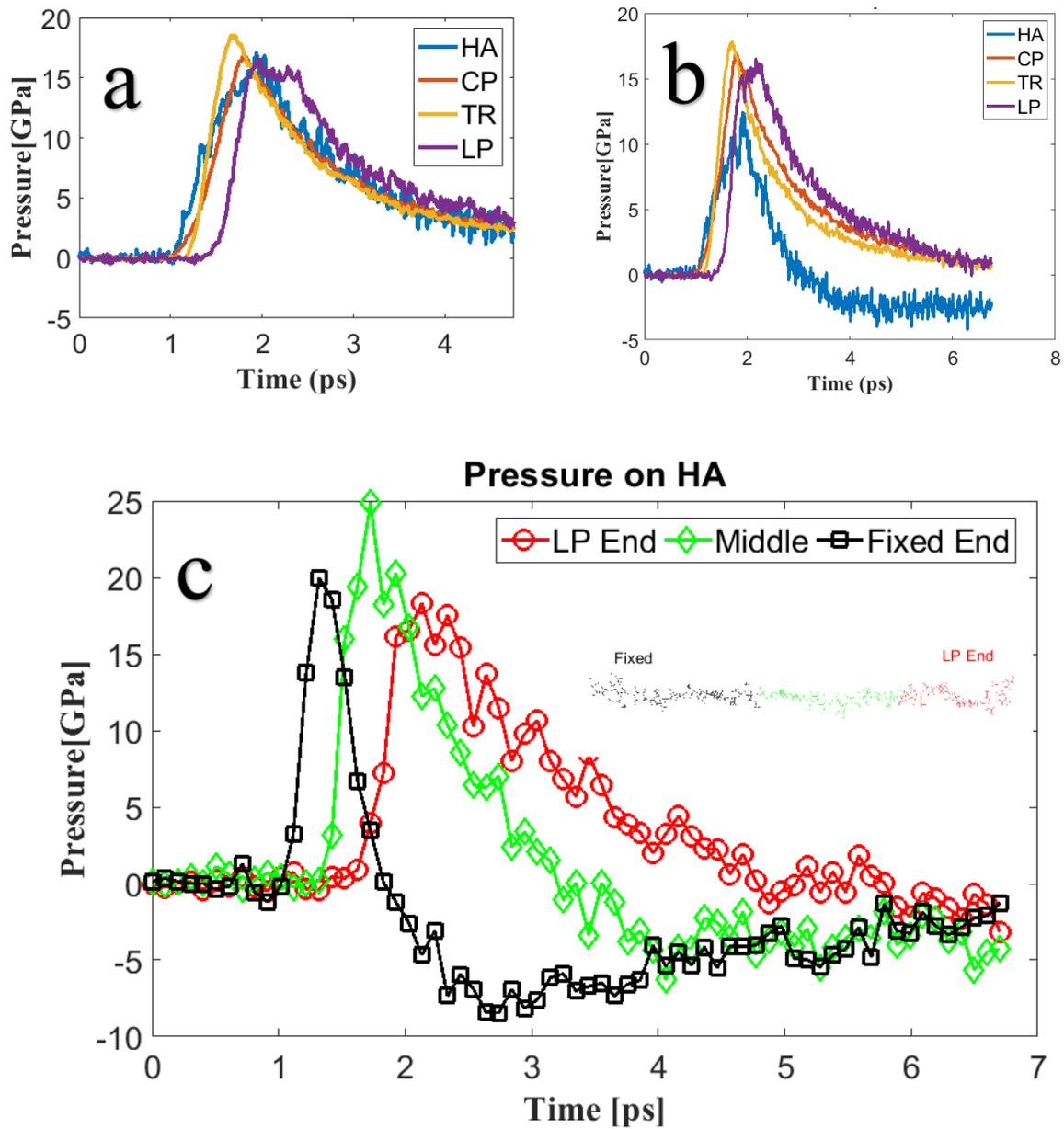

*Figure 10* Pressure profile on PNN components (a) Pressure on different PNN components in the absence of boundary condition (b) Pressure on various PNN components in the presence of boundary condition (c) Pressure at different location of HA

Finally, the number of hydrogen bonds has been measured at different conditions (*Figure 11*). As the shock speed is increasing, the number of hydrogen bonds severely impaired. A hydrogen bond is only found in the protein molecules, suggesting that the protein's secondary structure is damaged due to the shock wave. The lowest number of a hydrogen bond is found for the model with the preexisting bubble, suggesting that the bubble jet causes maximum damage to the PNN.

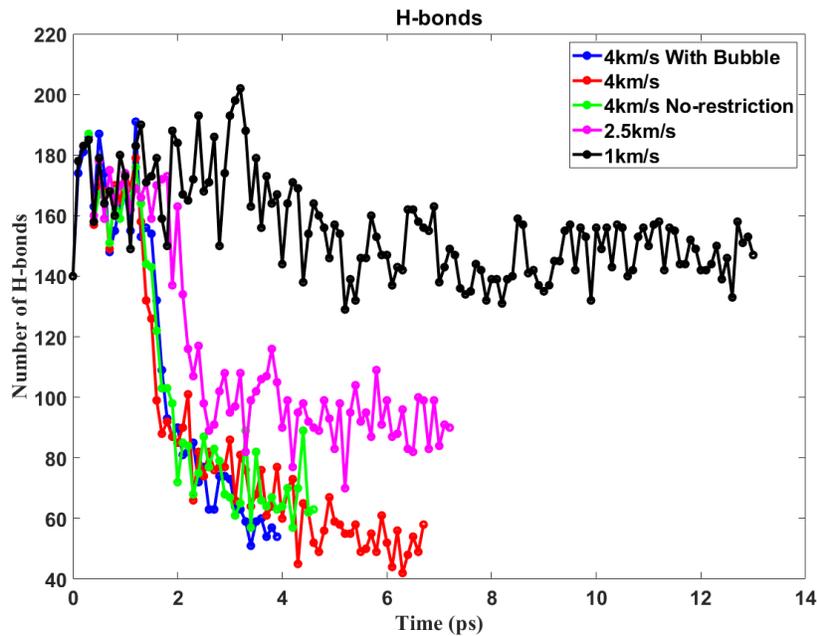

*Figure 11* Number of Hydrogen bonds at a different speed and boundary condition (cut off distance 3Å and angle 20º)

## 4. Conclusions:

The PNN network protects the neuron from physical damage, reduces oxidative stress, and conserve charge balance to facilitate neurotransmission. Therefore, the damage probability of PNN under shock loading needs to be evaluated. From the shock loading simulation, it can be concluded that:

- The protein structure is less prone to failure due to shock loading, while hyaluronan is the most vulnerable molecule to break during the shock loading.
- The damage efficiency is strongly dependent on the shock speed, presence of a bubble, and boundary condition. The presence of a bubble in the system initiates asymmetric collapse during shock propagation and exerts water jets to damage the molecules which are present in the projected domain.
- Although the protein components' pressure is still compressive, the significant reduction of the number of hydrogen bonds of the proteins makes it clear that the protein's secondary structure altered significantly at higher shock speed.

Although PNN does not have direct involvement in propagating the action potential from one neuron to another, the absence of PNN may cause severe disruption of neurotransmission at the synaptic cleft, thus leads to altered signal processing.

**Authors Contribution**

K.M. conducted all simulations and contributed to writing the manuscript. F.H. provided insights on PNN network strength. M.K. helps in simulating the biomolecules. A.A. helped in developing the concept. All authors reviewed the manuscript.

**Acknowledgment**

This work has been funded by the Computational Cellular Biology of Blast (C2B2) program through the Office of Naval Research (ONR) (Award #N00014-16-1-2142 and N00014-18-1-2082 - Dr. Timothy Bentley, Program Manager). The authors acknowledge the Texas Advanced Computing Center (TACC) at The University of Texas at Austin for providing HPC resources that have contributed to the research results reported within this paper.

URL: http://www.tacc.utexas.edu.

**Conflict of Interest**

The authors declare no conflict of interest.